# ABSTRACT

LIU, ZHUORAN. Interpreting Performance Profiles with Deep Learning. (Under the direction of Dr. Xu Liu).

Profiling tools (also known as profilers) play an important role in understanding program performance at runtime, such as hotspots, bottlenecks, and inefficiencies. While profilers have been proven to be useful, they give extra burden to software engineers. Software engineers, as the users, are responsible to interpret the complex performance data and identify actionable optimization in program source code. However, it can be challenging for users to associate inefficiencies with the program semantics, especially if the users are not the authors of the code, which limits the applicability of profilers.

In this thesis, we explore a new direction to combine performance profiles and program semantics with a deep learning approach. The key idea is to glean code summary for semantic information (at a certain level) and integrate it into a profiler, which can better understand program inefficiencies for actionable optimization. To be concrete, we combine profiles generated by Async Profiler (the state-of-the-art Java profiler) with code summarization from a fine-tuned CodeBERT-based model. We demonstrate the code summaries of any selected call path in a graphic user interface. Our system can effectively assist analysis on many Java benchmarks.

Interpreting Performance Profiles with Deep Learning

by
Zhuoran Liu

A dissertation submitted to the Graduate Faculty of
North Carolina State University
in partial fulfillment of the
requirements for the Degree of
Doctor of Philosophy

Computer Science

Raleigh, North Carolina
2022

APPROVED BY:

_________________________________  _________________________________
Dr. Xipeng Shen                     Dr. Guoliang Jin

_________________________________  _________________________________
Member 3 of Committee               Member 4 of Committee

_________________________________
Dr. Xu Liu
Chair of Advisory Committee

# **DEDICATION**

To my parents.

# BIOGRAPHY

The author was born in Shenyang, Liaoning, China. She completed her bachelor degree at Northeastern University in China. Since 2021 she has been pursuing her Master of Science degree in Computer Science at North Carolina State University, Raleigh, USA. She works as a Graduate Research Assistant under the guidance of Dr. Xu Liu.

# ACKNOWLEDGEMENTS

I would like to thank my advisor, Dr. Xu Liu, for his guidance, support, and patience. I would also like to thank all the professors for my courses to widen my view. I also would like to thank all my labmates for their help.

# TABLE OF CONTENTS

# LIST OF TABLES

# LIST OF FIGURES

# CHAPTER 1

# INTRODUCTION

The performance of computer systems and applications is critical in program development [Adhianto et al., 2010]. With increasing complexity and requirements, applications with better computation are expected to consume less time. Such complexity growth often reduces the efficiency of the entire software stack, resulting in wasted resources, diminished performance, and power consumption [Molyneaux, 2009]. Dynamic profiling tools,like OProfile [opr, 2018], perf [Linux, 2015], HPCToolkit [Adhianto et al., 2010], play a significant part in the program development process as they can record and explore the running process of complex applications. The dynamic analysis tools make it possible to analyze and identify potential issues that occur at runtime and affect the program's reliability. Such system or application bottlenecks must be determined by some performance tools which could assist developers, like performance bottlenecks, hot spots, and high memory usage. Based on reports from analysis tools, we can locate the issues, identify memory and performance problems, and then debug the applications. Through this process, parts of the code would gain overall speed up. The advantages of using this kind of tool are decreasing execution time and increasing resource utilization efficiency.

Moreover, the key is to understand and optimize the parts recognized as bottlenecks and hot spots associated with the results from these tools. The solutions to these issues at runtime identified by the dynamic code analysis tools require further analysis. For example, the investigation of the profile obtained by the program analysis tool can refer to hotspots, insufficient parallelism, and memory bottlenecks at runtime. In general, issues detected by profiling tools could be two categories: hotspot and inefficiency [Li et al., 2022]. Hotspot problems refer to resource usage such as space and time. Inefficiency issues refer to the waste of resource. For detected issues, developers need to analyze why those problems occur, like spending a long time or accessing extra memory. In this paper, we focus on the hotspot issues.

While such analysis results obtained by the profiling tools are generally valuable but are challenging to be widely used and adopted by developers. The reasons are: (1) profilers require users to spend a lot of time to be familiarized and researched; (2) the information provided is very abstract, not well integrated with the program, and not easy to comprehend; (3) even after the user locates the program problem, they still need to study the source code to optimize the program.

To address these challenges, some systems such as EasyView [eas, 2022], MagpieBridge [Luo et al., 2019], IBM AppScan [Inc.], Xanitizer [Pyt, 2022], try providing a graphic user inference for profiles in the interactive development environments (IDEs). These can display and visualize profiles, reducing abstraction.

Moreover, program analysis results need to be combined with an understanding of the source code to have more specific guidance during practical production. However, people who try to analyze and optimize the program usually differ from those who develop them. Despite this, they need to understand and interpret what the code illustrates and come up with clues about how to optimize them. Moreover, in the profile analysis stage, software engineers usually explore all the related parts of applications, which is difficult and time-consuming. Therefore, software engineers or developers often need to take full advantage of the practical information obtained by profiling tools.

Thus, high-quality descriptive documentation or code summaries are essential to this process. Software documentation, especially code comments, is vital in assisting developers in understanding the source code. According to statistics [Wang et al., 2022], programmers spend an average of 58% of their time understanding source code. Short descriptions offer programmers to understand more quickly what a piece of code does and what that code is used for among the whole program instead of spending more time reading the entire program itself. But sometimes, the program related to identified issues may not contain sufficient and accurate code comments. Moreover, documenting source code is a labor-consuming task [de Souza et al., 2005b; Kajko-Mattsson, 2005], and it is widespread for programmers to ignore writing some code comments while writing code [de Souza et al., 2005a; Kajko-Mattsson, 2005; Roehm et al., 2012; Shi et al., 2011]. The automatic or semi-automatic generation of code summaries is essential [Forward and Lethbridge, 2002]. There are numerous pre-trained models for programming language (PL) and natural language (NL) which can be applied into downstream code-to-text task, like CodeBERT [Feng et al., 2020], CoTexT [Phan et al., 2021], PLBART [Ahmad et al., 2021], ProphetNet-X [Qi et al., 2021].

Some systems support the visualization of profiles in popular integrated development environments for better analysis and visualization [Beck et al., 2013; Pangin, 2018]. Such solutions cannot solve the difficulty of interpreting profiling data only with visualization, not with a deeper understanding of source code in the coding environments. When an inefficient problem is detected in the profile and mapped back to the source code, developers want to understand the cause of the problem and then further optimize the program based on it. When issues found in profiles map to the source code, the corresponding code summary of the referred source code could support developers in understanding the code. At the same time, the code summary from software engineers or auto-generation is usually a sentence in a practical scenario, which is not enough to identify the key to the issues.

## 1.1 Contributions

To fill the gap between profiles and the scope according to the source code, we developed a VSCode extension to introduce a deep learning approach to this new scenario, interpreting profiles. Combined with the calling context at runtime, the code summary will better describe the meaning of the source code. The contributions of this work are as follows:

- We integrated profiles generated by Async-Profiler with code summarization from a fine-tuned CodeBERT-based model, a natural language model for PL to NL.

- We mapped hot functions identified by profiles to the source file to facilitate understanding the located source code. In this way, we can more clearly and quickly understand the cause of the inefficient problem and the location of the source code.

- We visualized code summaries of the highlighted calling context to address the profile information within the coding environments, displaying program semantics of any selected call path.

## 1.2  Thesis Organization

We organize this paper as follows. Chapter 2 discusses the related work. Chapter 3 overviews the workflow of the tool. Chapter 4 illustrates the implementation details. Chapter 5 shows evaluations on the tool. Chapter 6 shows some discussions on the tool. Chapter 7 demonstrates the conclusions.

# CHAPTER 2

# RELATED WORK

In this chapter, we review existing profilers for Java applications (shown in Section 2.1), programming language and natural language models (shown in Section 2.2), and systems with generating code comments (shown in Section 2.3).

## 2.1 Profilers for Java Applications

In developing Java programs, software engineers need to address not only bugs detected by the compiler but also performance issues existing in successful running programs, like memory usage and performance bottlenecks. Java profiling tools must be designed to monitor and locate these issues to improve the development of Java programs. Nowadays, many Java performance analysis strategies in the Java community can support developers in comprehending their program behaviors [Li et al., 2021]. Java performance analysis tools have two types: identifying runtime hotspots in CPU cycles or memory utilizations [opr, 2018; Corporation, 2018; ej-technologies GmbH, 2018; GmbH, 2018; Linux, 2015; Milenkovic et al., 2008], monitoring redundant calculations [Della Toffola et al., 2015; Dhok and Ramanathan,

2016; Nistor et al., 2013; Song and Lu, 2017].

VisualVM [Corporation, 2018], JProfiler [ej-technologies GmbH, 2018], and YourKit [GmbH, 2018] identify where execution hotspots of CPU or heap are, which are currently common and widely used by sampling. OS timers are applied as sampling engines to deliver periodic samples, leading to low overhead. However, the detected hotspots are not a critical factor affecting the efficiency of the code sometimes because the resources occupied by the hotspot may not be utilized or affect the primary running efficiency [Li et al., 2022]. It is possible that hot spots are not areas of inefficient code. Therefore, it is time-consuming for users to determine whether eliminating the identified hotspots can optimize the program's performance. The Java profilers applied bytecode instrumentation may cause some issues. The tool may lead to the overhead of stack trace collection. The tool may affect the accuracy of JIT optimization and escape analysis [author=Song, Linhai and Lu, Shan, 2019]. And the changes in code size may affect the compiled code arrangement and cache capability. And the instrumented bytecode may adjust the locations of safe-points [Nisbet et al., 2019; Mytkowicz et al., 2010].

Linux Perf [Linux, 2015], Async-profiler [Pangin, 2018], and Oprofile [opr, 2018] are based on hardware, applying PMU (Hardware Performance Monitoring Unit) to sample periodically. These PMU-based hotspot profilers supply better insights than OS timer-based hotspot profilers, as they categorize hotspots based on distinct performance metrics obtained by the PMU, such as cache misses and instruction count.

Our work is based on profiles from Async-Profiler, a famous and low-overhead profiler proposed by Google. Async-Profiler monitors CPU cycles, and memory allocations, and records hardware and Software performance counters such as cache misses, branch misses, page faults, context switches, and contented lock attempts.

## 2.2 Code Summarization Models

There are many automatic code summarization generation models. Code summarization refers to a transition from code to text, generating a natural language description automatically to assistant users understand the given code snippet. There are two approaches to generate code summary: one is to construct new sentences from the original code segment [Allamanis et al., 2016; Chen et al., 2019; Chen and Zhou, 2018; Hu et al., 2018a,b; Iyer et al., 2016; Wan et al., 2018], the other is to make use of information retrieval (IR) to search sentences from similar code segment as a summary [Eddy et al., 2013; Movshovitz-Attias and Cohen, 2013; Rodeghero et al., 2014; Wong et al., 2015; Haiduc et al., 2010]. These models are supported by large enough datasets to deal with a variety of scenarios.

Statistical models, including neural networks, frequently appear in code intelligence scenarios. Researchers applied many state-of-the-art pre-trained natural language processing (NLP) models to pre-trained models on the programming language (PL), like BERT [Devlin et al., 2018] and GPT [Solaiman et al., 2019]. It is indicated from the application of GLUE [Wang et al., 2018] for NLP and the usage of ImageNet [Deng et al., 2009] for computer vision that a diverse benchmark dataset has dramatically influenced the expansion of intelligence research. The attainment of pre-trained models for NLP guides the blooming of pre-trained models for PL, such as CodeBERT [Feng et al., 2020] and IntelliCode Compose [Svyatkovskiy et al., 2020], transformer-based [Vaswani et al., 2017] NLP models with PL datasets. CoTexT [Phan et al., 2021]is a pre-trained model supporting both "bimodal" and "unimodal" data. PLBART is a pre-trained model based on Java and Python programming language and natural language with the Denoising Autoencoders [Ahmad et al., 2021]. CodeBERT [Feng et al., 2020] generates the representations of PL on different pre-training tasks and equips these to support other downstream tasks, like code summarization and code classification.

There are other models to apply abstract syntax tree (AST) to assist them in learning PL. Jiang [Jiang et al., 2021] presents a tree-based pre-trained model for PL, combined with AST, called TreeBERT. TreeBERT takes the path set of AST as input, applies node position embedding, and then uses a mixed objective model to train the model. LeClair et al. [2020] uses both a series of source code tokens and ConvGNN to encode ASTs of Java methods and then generate a summary. Hu et al. [2018a] tagged sequences with information from the AST. Alon et al. [2018] used the paths extracted from the AST to help summarize. Meanwhile, Allamanis et al. [2017] introduced Graph Neural Networks (GNNs) to learn representations of codes. These works demonstrate the capability of neural networks to treat the form of a graph or a tree as its input to acquire knowledge from code and apply that to downstream tasks, like code summarization.

This paper decides to finetune the codeBERT model on the code-to-text task with Java language datasets, using the code's semantic-level information rather than the code's syntactic-level structure, like AST. The semantic-level information does not introduce excessive deep hierarchies, unlike AST, promoting the model performance.

## 2.3  Code Summary in the IDEs

As mentioned in the above section, deep learning models have matured on code-text tasks recently, enabling the emergence of systems that automatically generate code summaries in IDEs. Mintlify Doc Writer [Min, 2022] is a solution for generating auto-documentation, available at VScode and IntelliJ IDEA marketplace. Readable [rea, 2022] is also a plugin for generating comments, supporting methods and customized lines.

These existing systems provide comments inside the editor and automatically generate code documentation, including functional comments and property information, for highlighted code or code with the cursor. Our system combines automatic code comments with calling context [Zhao et al., 2020] to display a new view to allow users to analyze and optimize program problems effortlessly.

# CHAPTER 3

# OVERVIEW

This chapter overviews the structure of this presented system shown in Figure 3.1. Section 3.1 clarifies the features of Async Profiler which provides profiles for our system as input, Section 3.2 explains the support offered by EasyView as a platform, and Section 3.3 demonstrates CodeBERT and fine-tunes it to a new scenario.

Based on the EasyView system structure, our implemented parts are marked in yellow It mainly consists of 3 components: (1) a profiles converter (shown in Section 4.1) receives profiles from Async-Profiler and source code of the Java applications and interacts with the Data Abstract Interface provided by EasyView; (2) a code summary module is based on a fine-tuned CodeBERT model (shown in Section 4.2) and receives code segments from Data Analysis Interface provided by EasyView and passes the generated code summary to the Data Visualization Interface; (3) a code summary view (shown in Section 4.3) displays the code summary of the selected call path.

Our system accepts the user's click on the block on the flame graph view provided by EasyView as input, and the function corresponding to the clicked block is called the current node in this paper. Here are the following stages based on the input:

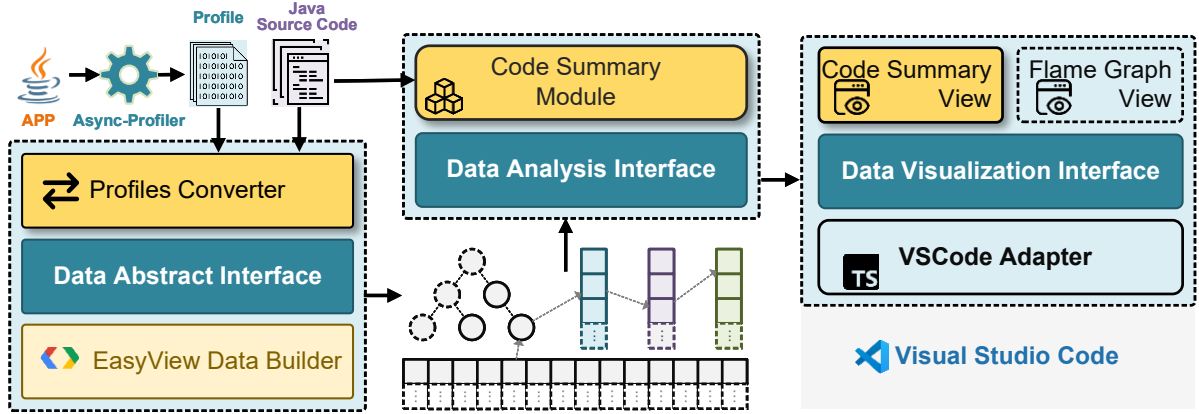

Figure 3.1: Overview of the infrastructure of our system. There are three components: profiles converter, code summary module, and code summary view.

1. We mapped the obtained nodes back to the file path and the specific line ranges of the source code based on the profiling data, provided by the profiles converter.

2. After receiving the function contents corresponding to these nodes based on the last step, input them into the deployed fine-tuned CodeBERT model.

3. Finally, the code summary module transmits the generated code summary back to the code summary view in the VSCode. The code summary view displays the summary of all nodes on the VSCode Webview in the form of a tree.

## 3.1 about Async-Profiler

Our system is based on the Async-Profiler profile [Pangin, 2018], an open-source sampling profiler for Java with low overhead. Async-Profiler is a popular Java analysis tool to detect HotSpot features and trace stack calls and memory allocations.

The Async-Profiler collects memory allocation information when analyzing heap allocations but does not use CPU-intensive code. Async-Profiler does not have a bytecode-based instrumentation strategy, which significantly impacts performance and has no effect on the accuracy of JIT optimization and Escape analysis.

When collecting stack trace samples, other technologies map call stacks from `perf_-events` with call stacks from `AsyncGetCallTrace` through an agent transform between addresses and Java method names. Additionally, Async-Profiler can maintain stack traces where `AsyncGetCallTrace` fails. As a result, Async-Profiler has the following advantages:

- Applicable for older Java versions and interpreter frameworks.

- Does not introduce significant performance overhead.

- Does not require the file mapping from Java code addresses to method names and the `perf.data` file for future use.

Additionally, if it is required to profile the java program at the beginning, Async-Profiler can be treated as a Java Agent.

Furthermore, Async-Profiler could output obtained performance metrics of the Java app to files in the format JFR. We could look at the profiling results at the terminal, which is not enough to understand deeply. Async-Profiler also provides APIs to convert metrics to files in other formats like `pprof` or functionality flame graph in HTML format. Intellij IDEA integrates Async-Profiler to profile running programs and view the flame graph [int, 2021]. These solutions do not provide different strategies to understand profiling results.

Our system applied Async-Profiler to profile data like method call information and memory allocation data and save the metrics as JFR files. And we convert profiling results to `pprof` files, which are readable in the EasyView [eas, 2022], providing a basis for a better understanding of profiles afterward.

## 3.2 about EasyView

EasyView is a VSCode extension to visualize profiling results available at VSCode market, supporting pprof format and various metrics. It provides top-down, bottom-up, flat views as flame graphs and tree tables. EasyView can associate flame graphs and tree tables with source code. Easyview visualizes all three forms of trees to drive better analysis supporting search. Show all three forms of trees as flame graphs:

- Top-down flame graph: the default view. The root represents the program entrance; the nodes under the root represent the call path; the length of each node represents the metric value.

- Bottom-up flame graph: inverting the top-down tree, showing the functions called in different call paths. It is helpful to judge where the hot function is being called from.

- Flat flame graph: showing loaded modules, functions, files, and functions as different hierarchies instead of call paths.

When we get the profile and the calling context tree from Async-Profiler, developers could get insights about hotspot and then study nodes (that is, methods) identified as hotspot with provided views. Moreover, we still need to know the meaning of each code snippet to figure out the optimization method. Users need to interpret the cause of the hotspot, so users usually need to read and understand more than a single function but also functions on the whole call path to assist in understanding the cause of the problem.

However, even if we can locate and view the source code referring to the identified hotspot with the visualization of the profiles, this is still a very time-consuming task. Therefore, we use the deep learning method to automatically generate the code summary to speed up the process of understanding the source code and make it probable for all users to fast-scan all the related functions, regardless of their knowledge of the source code. If we provide their code summary and combine this with the profile data, we could get a sentence or a paragraph to wrap up each node.

With the assistance of EasyView, we can view the flame graph based on pprof file from Async-Profiler mentioned in 3.1. With provided views, developers could get insights about HotSpot and then study nodes (that is, methods) identified as HotSpot. Our system provides more modules based on it. We maintained the tree structure to display the selected call path in profiles (that is, the highlighted call path), including the nodes (parent, current, and children nodes) as follows:

- All parent nodes of the current node till the roots (here refer to the highest in the range of available application code if top-down view, that is, callers). Parent nodes mean all the methods represented by these nodes in this paper;

- All children nodes of the current node (here refers to all methods called by the current node, that is, all nodes underneath in the top-down view, that is, callees). Children nodes mean all the methods represented by these nodes in this paper.

Take the top-down tree as an example, which shows the call paths by aggregating common callers. As shown in Figure 3.2, Assuming "foo" is a hot function (surrounded by an oval), then we provide code summarization for the selected call path (surrounded by dotted line): parent functions: "main", moo"; current function: "foo"; children function:

"print", "print", respectively. And combining profiles with the generated summary, we could understand the whole path more than a function.

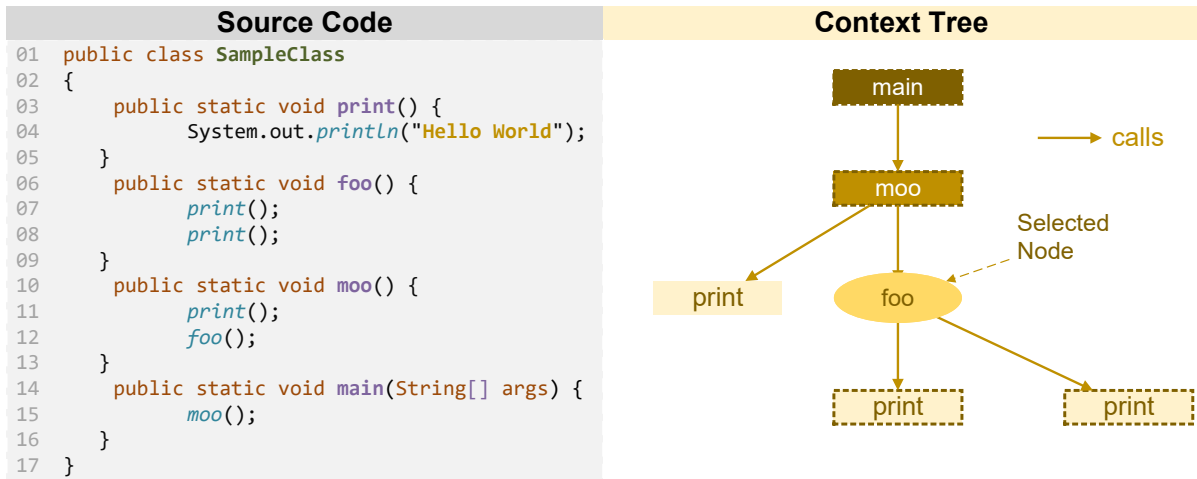

Figure 3.2: An example for the selected call path structure.

## 3.3 CodeBERT Model

We chose the CodeBERT model, a pre-trained model on NL-PL pairs in six programming languages (Python, Java, JavaScript, PHP, Ruby, Go) to do fine-tuning in 4.2. Our system applies the filtered CodeSearchNet dataset [Husain et al., 2019] to train the model, containing millions of pairs of methods and their docstrings for various programming languages. We will be focusing on the Java programming language in this paper. The experiments show CodeBERT obtains 17.65 as the BLEU score [Lin and Och, 2004] in Java language.

Given the call path of the application, users can locate hot functions manually. However, when they find the corresponding source lines, users still need to read the source code on their own, which is not sure how many lines are required to be read, not limited to the current function range. Our system applies CodeBERT to this scenario to assist in understanding profiling data. Code summaries generated by CodeBERT are displayed in VSCode as a tree in order of parent nodes, current nodes, and children nodes. The presentation enhances users' efficiency in analyzing the source code and accelerates the investigation of the cause of HotSpot. Based on the architecture of our software, we can replace CodeBERT with other PL-NL models as this part of the workflow. Similarly, our software is available to other profiling analysis tools in different programming languages.

# CHAPTER 4

# IMPLEMENTATION

This chapter outlines the implementation of this presented system. Section 4.1 illustrates how to convert the profiles for APIs of EasyView, Section 4.2 demonstrates the code summary module with a fine-tuned CodeBERT model, Section 4.3 shows how to support the code summary view.

## 4.1 Profiles Converter

Figure 4.1 shows the data mapping of our profiles converter. To construct the context tree, we merge all the common prefixes of the call paths. Combined with the first line of each Java file, we can find the source file corresponding to each context, which is the file path. Next, based on this mapping information, we can make each node of the context tree, which corresponds to metrics, now have a corresponding source file, function name, and code location. Finally, EasyView can map the nodes to the source code based on this information.

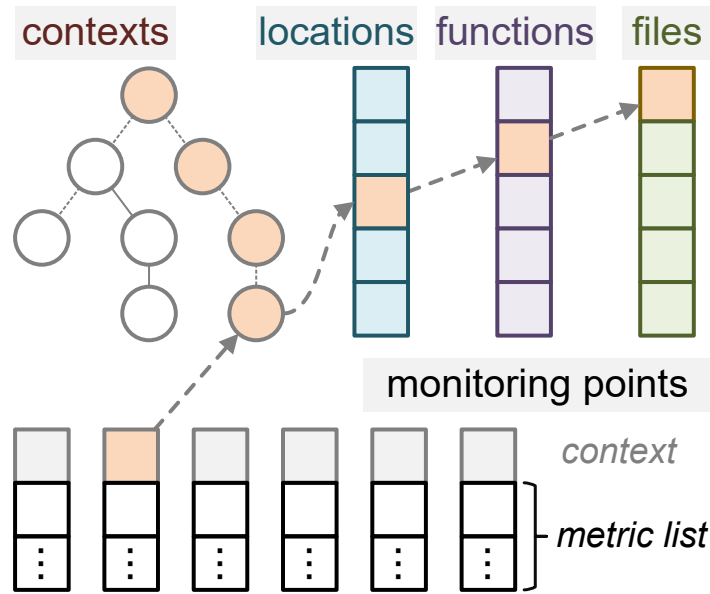

Figure 4.1: Profiles converter provides...

## 4.2 Code Summary Module

We use the CodeBERT pre-trained model to generate the code summary. To make it more suitable for this downstream task, we preprocess the original training dataset, which the CodeBERT pre-trained with, and fine-tune the CodeBERT with the code summarization task. Finally, to make the fine-tuned model available to VSCode, we deploy the model on the server side to inference the received code snippet, get a summary, and transmit it back to the client.

### 4.2.1 Data Cleaning

In this data-cleaning phase, we filter the dataset to improve dataset quality. We remove some pairs of functions and comments based on the following rules.

1. We remove the pairs if the number of comments tokens is less than four because those comments cannot contain valid information.

2. We remove the pairs if the pairs have non-ASCII characters because we only discuss English comments here.

3. We remove the links or tags starting with `http://` in comments, because we don't require the model to learn the external resources the documentation may contain.

4. We remove the pairs with the parameters that do not appear in JavaDoc's parameter list [jav, 2022] because such parameters are outdated.

5. We remove the pairs if comments are longer than the corresponding code, because such comments may not be code descriptions but usage or other text that is not related to the meaning of the code.

After this data cleaning, we then fine-tuned the CodeBERT model with this dataset.

### 4.2.2 Model Fine-tuning

In this fine-tuning part, we applied the Encoder-Decoder framework to solve the code summarization task, following Lu et al. [2021], shown in Figure 4.2. It utilizes CodeBERT as the encoder to generate vectors for the Decoder and a randomly initialized Transformer with six layers, 768- dimensional hidden states, and 12 attention heads as the Decoder. The max length of input and output are defined as 256 and 128, respectively. Finally, we operate the Adam optimizer to modify the parameters. The learning rate and the batch size for the fine-tuning are 5e-5 and 32, respectively.

To evaluate the models, we apply a smoothed BLEU score [Lin and Och, 2004] as an evaluation metric because this is suitable for evaluating short documents. The bigger the metric is, the more the generated results are close to natural languages. In this experiment, the fine-tuned model achieves (shown in Table 4.1) the state-of-the-art performance on six programming languages, superior to RoBERTa [Liu et al., 2019].

Accordingly, we have fine-tuned the CodeBERT, ready to do the inference task. We will apply the fine-tuned CodeBERT to inference the received functions and return code summary to the client.

## 4.3 Code Summary View

Viewing profiles with a flame graph may identify some hot functions based on memory allocations or CPU cycles. When users want to explore the hot function to interpret and

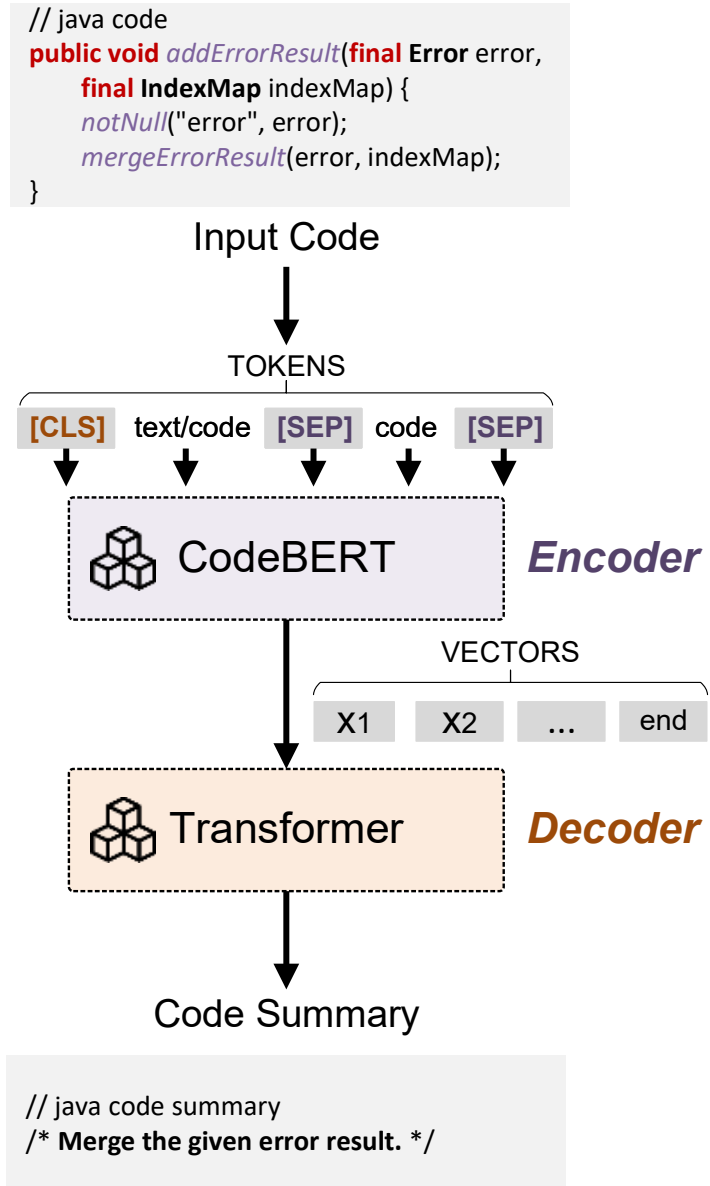

Figure 4.2: Overview of the Encoder-Decoder framework in our system. Input is the Java code segment and the output is the Java code summary.

Table 4.1: The BLEU scores achieved by different models on code summarization with Java Language.

| Model | BLEU Score |
|---|---|
| Seq2seq [Vaswani et al., 2017] | 15.09 |
| Transformer [Vaswani et al., 2017] | 16.26 |
| RoBERTa | 16.47 |
| CodeBERT | **17.65** |

optimize it, we zoom in on the flame graph based on the highlighted function. The next steps are as follows:

1. Based on the mapping method provided in Section 4.1, we trace back to the source file of the functions related to the selected hot function in the call path;

2. Then locate the whole function range through the line number in the respective source file;

3. Use regular expressions to process the located function code and send it to the server.

4. Next, we get the code summary from the server side.

We then show code summaries in a combination of vertical and horizontal tree structures, including all the callers of the selected function horizontally, all the callees of the function chosen vertically, and the selected function itself among them.

CHAPTER

EVALUATION

This chapter evaluated our system from two aspects: scalability and effectiveness. Scalability is based on whether we can add or improve existing functionality. Effectiveness is to explore how our system can provide optimization suggestions through the case study.

## 5.1 Extensibility

The Extensibility of this system is illustrated in two parts: an alternative deep learning method and a profiler with different programming languages.

1. Our system can adapt to different deep learning models. If the model accepts code snippets as input and can output code summary, it can replace the fine-tuned Code-BERT model.

2. Our system can substitute profiles from Async-Profiler with other proof format profiles. Because our system can encode proof format profiles and then use our converter to map the profiling data back to the source code, the implemented converter can map the input profile information back to the file path of the source code.

Such replacement operations do not influence the process of other modules because the modules are well-encapsulated.

## 5.2 Effectiveness

We evaluate the effectiveness by applying our system to Java application benchmarks to see how it boosts profiles. This section shows two case studies of our system, one is a micro-benchmark, and the other is from SPECjvm2008. They illustrate the insights provided by our system can help developers better facilitate profiles.

### 5.2.1 micro-benchmark

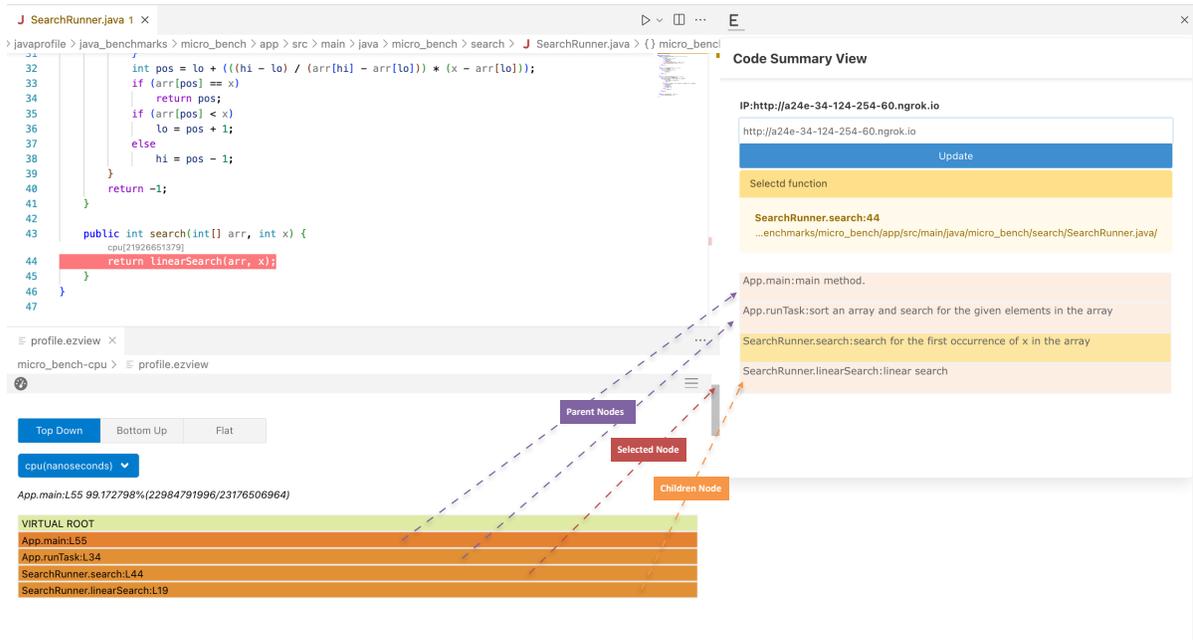

Figure 5.1: Our system identifies and assist to optimize the search method run time.

We first applied our system to an ideal benchmark. In this section, we choose a micro-benchmark shown in Figure 5.1. The flame graph shows the `search` method of `SearchRunner` consumes most CPU cycles of the total, and we cannot infer what the `runTask` of `App` does by the method name. So we look into the `search` method of `SearchRunner` to see how to optimize this hot function. Our system shows the `search` method of `SearchRunner` is called by the `runTask` method of `App` whose summary is: "sort an array and search for the given element in the array". Thus, we can understand that instead of linear search, we should use binary search on the given array in this part, which can speed up the program's running.

### 5.2.2 SPECjvm2008: Scimark.fft.large

We then applied our system to SPECjvm2008: `Scimark.fft.large`. Scimark [Sci, 2004] is a complex Java sub-benchmark of SPECjvm2008 to measure the performance of Java runtime environments. Scimark is a complex Java sub-benchmark of SPECjm2008 to measure the performance of Java runtime environments. We run `Scimark.fft` (a fast Fourier transform, FFT) with the large dataset provided by the benchmark.

When analyzing the case shown in Figure 5.2, we detect the `transform_internal` method of FFT as a hot function with the most counts of cache misses among the functions. Then we want to know why it happened, so we need to review its call path, concluding the nodes related to it from the lowest node to the highest node to find the reason.

We first study the callees of `transform_internal` method. Here our system offers summaries of this selected call path. We infer that the methods from `run` of `Benchmark` to the `run` of `FFT` are related to the benchmark execution. In this case, we could skip these methods to save time on nonrelevant codes. In the meantime, it shows a big gap between the length of `bitreverse` block and `transform_internal` block. We then review the `transform_internal` code and find that a 3-level loop nest is accessing array `data`. The loading stride for array `data` is large, leading to poor spatial locality. Thus, we interchange loop a and b to decrease the stride to optimize the program. This optimization reduces the 70%- of cache misses of the entire program, gaining a 2.37x speedup.

Figure 5.2: Our system identifies the array `data` with poor spatial localiity in the `transform_internal` method in SPECjvm2008: Scimark.fft.large.

# CHAPTER 6

# DISCUSSIONS

In this chapter, we discuss some limitations of our system.

First, General Language Understanding Evaluation benchmark for CODE (CodeXGLUE) Lu et al. [2021] tested some state-of-the-art models on Java code summarization task. There are three types of baseline models: (1) BERT-based pre-trained model like CodeBERT, which is suitable for understanding; (2) GPT-based pre-trained model, called CodeGPT, which supports completion and generation tasks; (3) Encoder-Decoder framework supports sequence-to-sequence generation tasks. Table 6.1 shows the BLEU scores achieved by different models such as DistillCodeT5, PolyglotCodeBERT [Ahmed and Devanbu, 2022], ProphetNet-X,CoTexT, PLBART, CodeBERT, RoBERTa, Transformer, Seq2Seq, on the code summarization with different programming languages, ranked in descending order. Thus, CodeBERT does not perform best on the code summarization task. Some models proposed later than it have higher scores, indicating that they can generate better code summary. Then, we can support other pre-trained models in the future.

Second, many performance analysis tools now support other programming languages besides Java, such as Python, Javascript, and Go. Correspondingly, many pre-trained models support the code summarization task in different programming languages. In this way, we can support other programming languages. Lastly, we only deliver extensions for VSCode now, while developers also utilize other IDEs, like IntelliJ IDEA, Goland, and Clion. We could then release plugins for other IDEs.

Table 6.1: The BLEU scores achieved by different models on code summarization.

| Model | Java | Ruby | JavaScrtipt | Go | PHP | Python |
|---|---|---|---|---|---|---|
| PolyglotCodeBERT | **20.11** | **14.75** | **15.80** | 18.77 | **26.23** | 18.71 |
| ProphetNet-X | 19.39 | 14.37 | 16.6 | 18.43 | 24.57 | 17.87 |
| CoTexT | 19.06 | 14.02 | 14.96 | 18.86 | 24.68 | **19.73** |
| PLBART | 18.45 | 14.11 | 15.56 | **18.91** | 23.58 | 19.3 |
| CodeBERT | 17.65 | 12.16 | 14.90 | 18.07 | 25.16 | 19.06 |
| RoBERTa | 16.47 | 11.17 | 11.90 | 17.72 | 24.02 | 18.14 |
| Transformer | 16.26 | 11.18 | 11.59 | 16.38 | 22.12 | 15.81 |
| Seq2Seq | 15.09 | 9.64 | 10.21 | 13.98 | 21.08 | 15.93 |

# CHAPTER 7

# CONCLUSIONS

## 7.1 Overarching Conclusions

In this thesis, we attempted to combine performance profiles with a pre-trained deep learning for PL to NL to infer program semantics. The main target is to display the collected code summary in profilers, to interpret and optimize program inefficiencies. To be specific, we use the profiles provided by Async-Profiler with code summary from a fine-tuned CodeBERT-based model. In addition, we deliver a graphic user interface to the Code summary of the highlighted call path. The automatic code summarization is in combination with the calling context information for the first time, which makes it easier to expose inefficiencies to users. Our system shows the capability to offer insights into many Java benchmarks.

## 7.2 Future Work

We preview some future directions. First, we plan to release our system to other popular IDEs, such as IntelliJ, IDEA. Second, to adapt to more programming languages, like Python and Go, with different profilers. Third, we plan to improve the deep learning model by adding the call path into the model's input to improve the code summary. It will be more information for the model to learn and to assist in interpreting hot functions that we combine the call path information with source code rather than simply showing a series of code summaries in the call path.

# REFERENCES

Scimark. https://www.spec.org/jvm2008/docs/benchmarks/scimark.html, 2004.

Oprofiler. https://oprofile.sourceforge.io/news/, 2018.

Get started with profiling in intellij idea. https://blog.jetbrains.com/idea/tag/async-profiler/, 2021.

Mintlify doc writer. https://mintlify.com/, 2022.

The Python Profilers. https://docs.python.org/3/library/profile.html, 2022.

Easyview. https://marketplace.visualstudio.com/items?itemName=xuhpclib-easyview.easyview, 2022.

Javadoc. https://www.oracle.com/java/technologies/javase/javadoc.html, 2022.

Readable. https://readable.so/, 2022.

Laksono Adhianto, Sinchan Banerjee, Mike Fagan, Mark Krentel, Gabriel Marin, John Mellor-Crummey, and Nathan R Tallent. Hpctoolkit: Tools for performance analysis of optimized parallel programs. *Concurrency and Computation: Practice and Experience*, 22(6):685–701, 2010.

Wasi Uddin Ahmad, Saikat Chakraborty, Baishakhi Ray, and Kai-Wei Chang. Unified pre-training for program understanding and generation, 2021. URL https://arxiv.org/abs/2103.06333.

Toufique Ahmed and Premkumar Devanbu. Multilingual training for software engineering. In *Proceedings of the 44th International Conference on Software Engineering*. ACM, may 2022. doi: 10.1145/3510003.3510049. URL https://doi.org/10.1145%2F3510003.3510049.

Miltiadis Allamanis, Hao Peng, and Charles Sutton. A convolutional attention network for extreme summarization of source code. In *International conference on machine learning*, pages 2091–2100. PMLR, 2016.

Miltiadis Allamanis, Marc Brockschmidt, and Mahmoud Khademi. Learning to represent programs with graphs. *arXiv preprint arXiv:1711.00740*, 2017.

Uri Alon, Shaked Brody, Omer Levy, and Eran Yahav. code2seq: Generating sequences from structured representations of code. *arXiv preprint arXiv:1808.01400*, 2018.

authors author=Song, Linhai and Lu, Shan. Roaringbitmap. `https://github.com/RoaringBitmap/RoaringBitmap`, 2019.

Fabian Beck, Oliver Moseler, Stephan Diehl, and Günter Daniel Rey. In situ understanding of performance bottlenecks through visually augmented code. In *2013 21st International Conference on Program Comprehension (ICPC)*, pages 63–72, 2013. doi: 10.1109/ICPC.2013.6613834.

Qingying Chen and Minghui Zhou. A neural framework for retrieval and summarization of source code. In *2018 33rd IEEE/ACM International Conference on Automated Software Engineering (ASE)*, pages 826–831. IEEE, 2018.

Qiuyuan Chen, Han Hu, and Zhaoyi Liu. Code summarization with abstract syntax tree. In *International Conference on Neural Information Processing*, pages 652–660. Springer, 2019.

Oracle Corporation. All-in-one java troubleshooting tool. `https://visualvm.github.io/`, 2018.

Sergio Cozzetti B. de Souza, Nicolas Anquetil, and Káthia M. de Oliveira. A study of the documentation essential to software maintenance. In *Proceedings of the 23rd Annual International Conference on Design of Communication: Documenting amp; Designing for Pervasive Information*, SIGDOC '05, page 68–75, New York, NY, USA, 2005a. Association for Computing Machinery. ISBN 1595931759. doi: 10.1145/1085313.1085331. URL `https://doi.org/10.1145/1085313.1085331`.

Sergio Cozzetti B de Souza, Nicolas Anquetil, and Káthia M de Oliveira. A study of the documentation essential to software maintenance. In *Proceedings of the 23rd annual international conference on Design of communication: documenting & designing for pervasive information*, pages 68–75, 2005b.

Luca Della Toffola, Michael Pradel, and Thomas R Gross. Performance problems you can fix: A dynamic analysis of memoization opportunities. *ACM SIGPLAN Notices*, 50(10):607–622, 2015.

Jia Deng, Wei Dong, Richard Socher, Li-Jia Li, Kai Li, and Li Fei-Fei. Imagenet: A large-scale hierarchical image database. In *2009 IEEE conference on computer vision and pattern recognition*, pages 248–255. Ieee, 2009.

Jacob Devlin, Ming-Wei Chang, Kenton Lee, and Kristina Toutanova. Bert: Pre-training of deep bidirectional transformers for language understanding. *arXiv preprint arXiv:1810.04805*, 2018.

Monika Dhok and Murali Krishna Ramanathan. Directed test generation to detect loop inefficiencies. In *Proceedings of the 2016 24th ACM SIGSOFT International Symposium on Foundations of Software Engineering*, pages 895–907, 2016.

Brian P Eddy, Jeffrey A Robinson, Nicholas A Kraft, and Jeffrey C Carver. Evaluating source code summarization techniques: Replication and expansion. In *2013 21st International Conference on Program Comprehension (ICPC)*, pages 13–22. IEEE, 2013.

ej-technologies GmbH. The award-winning all-in-one java profiler. https://www.ej-technologies.com/products/jprofiler/overview.html, 2018.

Zhangyin Feng, Daya Guo, Duyu Tang, Nan Duan, Xiaocheng Feng, Ming Gong, Linjun Shou, Bing Qin, Ting Liu, Daxin Jiang, and Ming Zhou. Codebert: A pre-trained model for programming and natural languages, 2020. URL https://arxiv.org/abs/2002.08155.

Andrew Forward and Timothy C. Lethbridge. The relevance of software documentation, tools and technologies: A survey. In *Proceedings of the 2002 ACM Symposium on Document Engineering*, DocEng '02, page 26–33, New York, NY, USA, 2002. Association for Computing Machinery. ISBN 1581135947. doi: 10.1145/585058.585065. URL https://doi.org/10.1145/585058.585065.

YourKit GmbH. The industry leader in .net java profiling. https://www.yourkit.com/, 2018.

Sonia Haiduc, Jairo Aponte, Laura Moreno, and Andrian Marcus. On the use of automated text summarization techniques for summarizing source code. In *2010 17th Working Conference on Reverse Engineering*, pages 35–44. IEEE, 2010.

Xing Hu, Ge Li, Xin Xia, David Lo, and Zhi Jin. Deep code comment generation. In *2018 IEEE/ACM 26th International Conference on Program Comprehension (ICPC)*, pages 200–20010. IEEE, 2018a.

Xing Hu, Ge Li, Xin Xia, David Lo, Shuai Lu, and Zhi Jin. Summarizing source code with transferred api knowledge. 2018b.

Hamel Husain, Ho-Hsiang Wu, Tiferet Gazit, Miltiadis Allamanis, and Marc Brockschmidt. Codesearchnet challenge: Evaluating the state of semantic code search, 2019. URL https://arxiv.org/abs/1909.09436.

IBM Inc. IBM AppScan. https://www.ibm.com/security.

Srinivasan Iyer, Ioannis Konstas, Alvin Cheung, and Luke Zettlemoyer. Summarizing source code using a neural attention model. In *Proceedings of the 54th Annual Meeting of the Association for Computational Linguistics (Volume 1: Long Papers)*, pages 2073–2083, 2016.

Xue Jiang, Zhuoran Zheng, Chen Lyu, Liang Li, and Lei Lyu. Treebert: A tree-based pre-trained model for programming language. In *Uncertainty in Artificial Intelligence*, pages 54–63. PMLR, 2021.

Mira Kajko-Mattsson. A survey of documentation practice within corrective maintenance. *Empirical Software Engineering*, 10(1):31–55, 2005.

Alexander LeClair, Sakib Haque, Lingfei Wu, and Collin McMillan. Improved code summarization via a graph neural network. In *Proceedings of the 28th international conference on program comprehension*, pages 184–195, 2020.

Bolun Li, Pengfei Su, Milind Chabbi, Shuyin Jiao, and Xu Liu. Djxperf: Identifying memory inefficiencies via object-centric profiling for java, 2021. URL https://arxiv.org/abs/2104.03388.

Bolun Li, Hao Xu, Qidong Zhao, Pengfei Su, Milind Chabbi, Shuyin Jiao, and Xu Liu. OJXPerf. In *Proceedings of the 44th International Conference on Software Engineering*. ACM, may 2022. doi: 10.1145/3510003.3510083. URL https://doi.org/10.1145%2F3510003.3510083.

Chin-Yew Lin and Franz Josef Och. Orange: a method for evaluating automatic evaluation metrics for machine translation. In *COLING 2004: Proceedings of the 20th International Conference on Computational Linguistics*, pages 501–507, 2004.

Linux. Linux perf tool. https://www.brendangregg.com/perf.html, 2015.

Yinhan Liu, Myle Ott, Naman Goyal, Jingfei Du, Mandar Joshi, Danqi Chen, Omer Levy, Mike Lewis, Luke Zettlemoyer, and Veselin Stoyanov. Roberta: A robustly optimized bert pretraining approach. *arXiv preprint arXiv:1907.11692*, 2019.

Shuai Lu, Daya Guo, Shuo Ren, Junjie Huang, Alexey Svyatkovskiy, Ambrosio Blanco, Colin B. Clement, Dawn Drain, Daxin Jiang, Duyu Tang, Ge Li, Lidong Zhou, Linjun Shou, Long Zhou, Michele Tufano, Ming Gong, Ming Zhou, Nan Duan, Neel Sundaresan, Shao Kun Deng, Shengyu Fu, and Shujie Liu. Codexglue: A machine learning benchmark dataset for code understanding and generation. *CoRR*, abs/2102.04664, 2021.

Linghui Luo, Julian Dolby, and Eric Bodden. MagpieBridge: A General Approach to Integrating Static Analyses into IDEs and Editors (Tool Insights Paper). In Alastair F. Donaldson, editor, *33rd European Conference on Object-Oriented Programming (ECOOP 2019)*, volume 134 of *Leibniz International Proceedings in Informatics (LIPIcs)*, pages 21:1–21:25, Dagstuhl, Germany, 2019. Schloss Dagstuhl–Leibniz-Zentrum fuer Informatik. ISBN 978-3-95977-111-5. doi: 10.4230/LIPIcs.ECOOP.2019.21. URL http://drops.dagstuhl.de/opus/volltexte/2019/10813.

Milena Milenkovic, Scott Jones, Frank Levine, and Enio Pineda. Performance inspector tools with instruction tracing and per-thread/function profiling. In *Linux Symposium*, 2008.

Ian Molyneaux. *The Art of Application Performance Testing: Help for Programmers and Quality Assurance*. O'Reilly Media, Inc., 1st edition, 2009. ISBN 0596520662, 9780596520663.

Dana Movshovitz-Attias and William Cohen. Natural language models for predicting programming comments. In *Proceedings of the 51st Annual Meeting of the Association for Computational Linguistics (Volume 2: Short Papers)*, pages 35–40, 2013.

Todd Mytkowicz, Amer Diwan, Matthias Hauswirth, and Peter F. Sweeney. Evaluating the accuracy of java profilers. *SIGPLAN Not.*, 45(6):187–197, jun 2010. ISSN 0362-1340. doi: 10.1145/1809028.1806618. URL https://doi.org/10.1145/1809028.1806618.

Andy Nisbet, Nuno Miguel Nobre, Graham Riley, and Mikel Luján. Profiling and tracing support for java applications. In *Proceedings of the 2019 ACM/SPEC International Conference on Performance Engineering*, pages 119–126, 2019.

Adrian Nistor, Linhai Song, Darko Marinov, and Shan Lu. Toddler: Detecting performance problems via similar memory-access patterns. In *2013 35th International Conference on Software Engineering (ICSE)*, pages 562–571. IEEE, 2013.

Andre Pangin. Async profiler. https://github.com/jvm-profiling-tools/async-profiler, 2018.

Long Phan, Hieu Tran, Daniel Le, Hieu Nguyen, James Annibal, Alec Peltekian, and Yanfang Ye. CoTexT: Multi-task learning with code-text transformer. In *Proceedings of the 1st Workshop on Natural Language Processing for Programming (NLP4Prog 2021)*, pages 40–47, Online, August 2021. Association for Computational Linguistics. doi: 10.18653/v1/2021.nlp4prog-1.5. URL https://aclanthology.org/2021.nlp4prog-1.5.

Weizhen Qi, Yeyun Gong, Yu Yan, Can Xu, Bolun Yao, Bartuer Zhou, Biao Cheng, Daxin Jiang, Jiusheng Chen, Ruofei Zhang, et al. Prophetnet-x: Large-scale pre-training models for english, chinese, multi-lingual, dialog, and code generation. *arXiv preprint arXiv:2104.08006*, 2021.

Paige Rodeghero, Collin McMillan, Paul W McBurney, Nigel Bosch, and Sidney D'Mello. Improving automated source code summarization via an eye-tracking study of programmers. In *Proceedings of the 36th international conference on Software engineering*, pages 390–401, 2014.

Tobias Roehm, Rebecca Tiarks, Rainer Koschke, and Walid Maalej. How do professional developers comprehend software? In *Proceedings of the 34th International Conference on Software Engineering*, ICSE '12, page 255–265. IEEE Press, 2012. ISBN 9781467310673.

Lin Shi, Hao Zhong, Tao Xie, and Mingshu Li. An empirical study on evolution of api documentation. In *Proceedings of the 14th International Conference on Fundamental Approaches to Software Engineering: Part of the Joint European Conferences on Theory* *and Practice of Software*, FASE'11/ETAPS'11, page 416–431, Berlin, Heidelberg, 2011. Springer-Verlag. ISBN 9783642198106.

Irene Solaiman, Miles Brundage, Jack Clark, Amanda Askell, Ariel Herbert-Voss, Jeff Wu, Alec Radford, Gretchen Krueger, Jong Wook Kim, Sarah Kreps, et al. Release strategies and the social impacts of language models. *arXiv preprint arXiv:1908.09203*, 2019.

Linhai Song and Shan Lu. Performance diagnosis for inefficient loops. In *2017 IEEE/ACM 39th International Conference on Software Engineering (ICSE)*, pages 370–380. IEEE, 2017.

Alexey Svyatkovskiy, Shao Kun Deng, Shengyu Fu, and Neel Sundaresan. Intellicode compose: Code generation using transformer. In *Proceedings of the 28th ACM Joint Meeting on European Software Engineering Conference and Symposium on the Foundations of Software Engineering*, pages 1433–1443, 2020.

Ashish Vaswani, Noam Shazeer, Niki Parmar, Jakob Uszkoreit, Llion Jones, Aidan N Gomez, Łukasz Kaiser, and Illia Polosukhin. Attention is all you need. *Advances in neural information processing systems*, 30, 2017.

Yao Wan, Zhou Zhao, Min Yang, Guandong Xu, Haochao Ying, Jian Wu, and Philip S Yu. Improving automatic source code summarization via deep reinforcement learning. In *Proceedings of the 33rd ACM/IEEE international conference on automated software engineering*, pages 397–407, 2018.

Alex Wang, Amanpreet Singh, Julian Michael, Felix Hill, Omer Levy, and Samuel R Bowman. Glue: A multi-task benchmark and analysis platform for natural language understanding. *arXiv preprint arXiv:1804.07461*, 2018.

Wenhua Wang, Yuqun Zhang, Yulei Sui, Yao Wan, Zhou Zhao, Jian Wu, Philip S. Yu, and Guandong Xu. Reinforcement-learning-guided source code summarization using hierarchical attention. *IEEE Transactions on Software Engineering*, 48(1):102–119, 2022. doi: 10.1109/TSE.2020.2979701.

Edmund Wong, Taiyue Liu, and Lin Tan. Clocom: Mining existing source code for automatic comment generation. In *2015 IEEE 22nd International Conference on Software Analysis, Evolution, and Reengineering (SANER)*, pages 380–389. IEEE, 2015.

Qidong Zhao, Xu Liu, and Milind Chabbi. Drcctprof: A fine-grained call path profiler for arm-based clusters. In *SC20: International Conference for High Performance Computing, Networking, Storage and Analysis*, pages 1–16. IEEE, 2020.